# Optical Writing of Magnetic Properties by Remanent Photostriction

V. Iurchuk,[1] D. Schick,[2] J. Bran,[1] D. Colson,[3] A. Forget,[3] D. Halley,[1] A. Koc,[2,4] M. Reinhardt,[2,4] C. Kwamen,[2,4] N. A. Morley,[5] M. Bargheer,[2,4] M. Viret,[3] R. Gumeniuk,[6] G. Schmerber,[1] B. Doudin,[1] and B. Kundys[1,*]

[1]*Institut de Physique et Chimie des Matériaux de Strasbourg (IPCMS), UMR 7504 CNRS-UdS 23 rue du Loess, 67034 Cedex 2, Strasbourg, France*
[2]*Institute for Methods and Instrumentation for Synchrotron Radiation Research, Helmholtz-Zentrum Berlin, Albert-Einstein-Straße 15, 12489 Berlin, Germany*
[3]*SPEC, CEA, CNRS, Université Paris, Saclay, CEA Saclay, 91191 Gif sur Yvette, France*
[4]*Institut für Physik & Astronomie, Universität Potsdam, Karl-Liebknecht-Straße 24-25, 14476 Potsdam/Golm, Germany*
[5]*University of Sheffield, Department of Materials Science and Engineering, Mappin Street, Sheffield S1 3JD, United Kingdom*
[6]*Institut für Experimentelle Physik, TU Bergakademie Freiberg, Leipziger Straße 23, 09596 Freiberg, Germany*

We present an optically induced remanent photostriction in $BiFeO_3$, resulting from the photovoltaic effect, which is used to modify the ferromagnetism of Ni film in a hybrid $BiFeO_3$/Ni structure. The 75% change in coercivity in the Ni film is achieved via optical and nonvolatile control. This photoferromagnetic effect can be reversed by static or ac electric depolarization of $BiFeO_3$. Hence, the strain dependent changes in magnetic properties are written optically, and erased electrically. Light-mediated straintronics is therefore a possible approach for low-power multistate control of magnetic elements relevant for memory and spintronic applications.

Multiferroic phenomena are often summarized in a Venn diagram showing the intersection of ferromagnetic, ferroelectric, and ferroelastic orders [1], each with its own control field. Numerous electric methods of magnetization control use elastic strain to leverage magnetoelectric (ME) properties in solids [2–4] and in magnetostrictive-electrostrictive structures [5–8]. The expected technological benefit is the possibility of low-power [9–11] operation down to the nanoscale [12–15]. Indeed, strain-mediated electric control of magnetic performance of tunnel junctions has been reported [15]. Furthermore, by using the ferroelastic effect of remanent strain, multiple nonvolatile states can be written on piezoelectric substrates [16–18]. Here we present the optical analog of this memory imprint approach, based on photostriction in $BiFeO_3$ (BFO) [19], a well-studied benchmark multiferroic material [20] exhibiting cross-linked ferroic orders. Light brings a new layer of functionality to multiferroics [21–24]. In particular, photoferroelectric [25] effects associated with above-band gap photovoltaic (PV) properties [26–28] can mediate light-induced changes of the ferroelastic order. While it is increasingly well established that BFO exhibits strain under illumination [29–31], the possibility of remanent strain states suggests a new approach [32]. The optical control of strain is particularly important for BFO, which possesses both high photostrictive efficiency [32] and large optoelastic coupling [33]. Furthermore, the magnetoelastic coupling in BFO has been shown to dominate its ME properties [34] that can provide a bridge for ME coupling between magnetic and electric orders [35]. These effects, together with the strain-tunable magnonic response in BFO thin films [36] provide an attractive strain-engineering prospective [37]. Photostriction control can also be extended to miniaturized structures using light-polarization-dependent functionality in ferroelectric domain walls in $BaTiO_3$ [38] offering an optical degree of control in spin-based devices [39,40]. Here we will first show that light can impact the internal electric field of BFO through the PV effect to produce optically induced ferroelastic remanent states, and then demonstrate the use of this ferroelastic deformation to stress a superposed ferromagnetic film, thereby achieving strain-mediated optical control of the magnetic anisotropy.

Illuminating a material which is ferroelectric (FE) and PV results in above-band gap voltage generation that changes the internal electric field in the sample [41]. The former process can be compared to the action of "subcoercive" electric fields insufficient to saturate the polarization, resulting in minor (nonswitching) FE loops [42]. Figure 1(a) illustrates how light excitation can be an alternative to the electric field, and generate a minor remanent polarization state via the PV effect [Fig. 1(b)]. A continuous wave (cw) 404 nm laser with a 3 ns rise time was used as the illumination source through an optical fiber. The sample was illuminated through a thin (20 nm) Au film, used as contact semi-transparent electrode for depolarizing the substrate. Under constant illumination, a steady-state photocurrent results in an increase of polarization saturating after ∼70 sec (not shown). The light induced change in electric polarization partly persists in ∼5.5% after the light is switched off [Fig. 1(a)]. One can conclude from Fig. 1(a) that different remanent polarization levels can result from different illumination times. The electric polarization of the BFO is the primary order parameter and it results in a change in strain (which is the secondary order parameter) that is linearly related to the polarization in the subcoercive region through the piezoelectric response of the oxide [43]. Figure 1(c) shows the remanent photostriction detected using a resistive measurement of a Ni thin film adlayer in the setup illustrated in Fig. 1(d). The overall remanent strain of the sample in Fig. 1(d) is tensile in the (010) plane and results in an in-plane expansion of the Ni film. In order to verify the remanent deformation of the BFO substrate, we carried out static x-ray diffraction experiments [Fig. 1(e)] at the XPP/KMC3 beam line in the synchrotron facility BESSY II (Berlin, Germany) [44]. A similar BFO crystal with the same orientation (but without adlayer) was used to determine the lattice spacing along the [010] direction in the as-grown [45] state and after 3 sec of light illumination. In this case, a femtosecond pulsed laser was used yielding a similar integral number of photons to that used for the switching in Figs. 1(a)–1(c) with the cw laser.

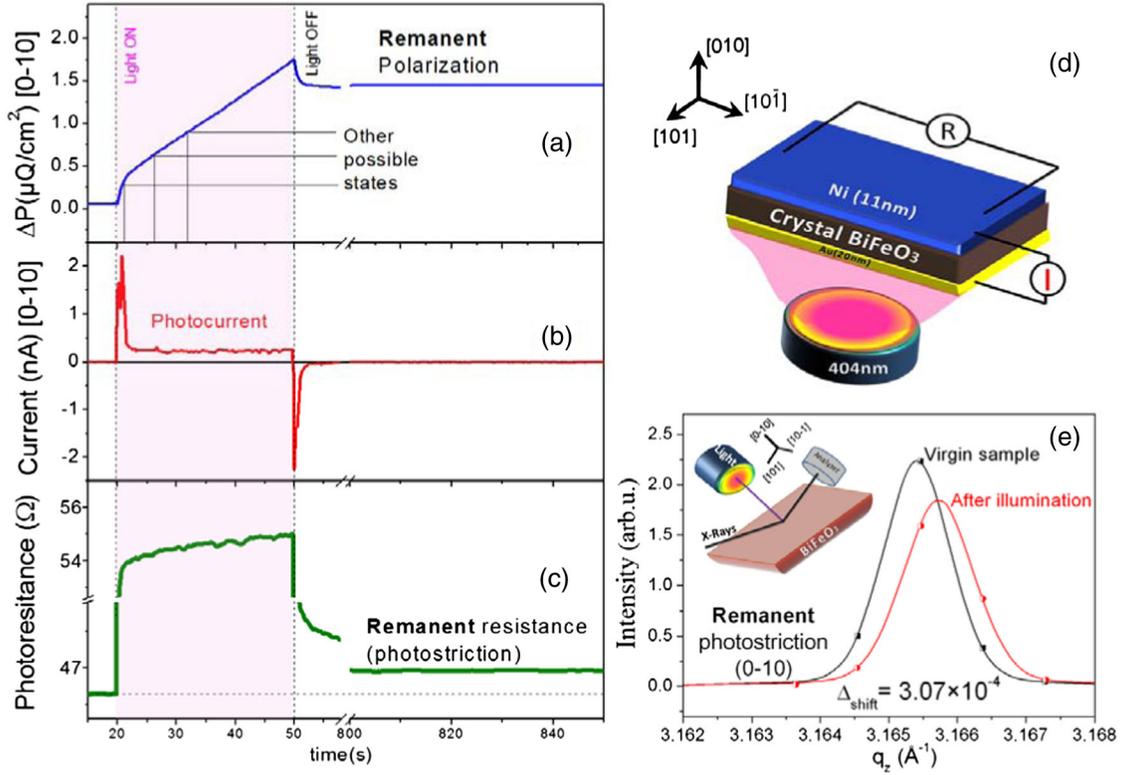

FIG. 1. (a) The remanent polarization state created by 30 sec UV light. (b) Corresponding photocurrent. (c) Remanent photostriction detected by measuring the resistance of the Ni film (d) and by static x-ray diffraction of BFO (e).

The pulsed laser consists of a multistage oscillator and amplifier system (Impulse, Clark-MXR) and delivers 250 fs long pulses of 10 $\mu$J pulse energy at a central wavelength of 1030 nm and a repetition rate of 208.3 kHz. They are then passed through a third harmonic setup at the beam line to generate the laser pump pulses of 350 nm with a final average power of 80 mW incident on the sample in a spot size of $277 \times 176$ $\mu m^2$ (FWHM) under an incidence angle of 20° between laser beam and sample surface. The x-ray photon energy was set to 9 keV with a relative bandwidth of $\Delta E/E = 10^{-3}$. The x-ray spot size on the sample was approximately 100 $\mu m^2$ and the experiment was conducted on a 4-axis goniometer in $\theta/2\theta$ geometry, with the diffracted photons detected by a DECTRIS Pilatus 100k hybrid-pixel 2D detector.

After illumination, the x-ray scan reveals a remanent shift of $\Delta q = 3.07 \times 10^{-4}$ Å$^{-1}$, which corresponds to a relative lattice contraction of $1 \times 10^{-4}$ along [010]. It is accompanied by a peak broadening in the out- and in-plane directions, which may be attributed to increase of intrinsic nanoscale inhomogeneities, possibly related to ferroelastic domains. No significant sample heating is expected during the x-ray scan as this would yield lattice expansion, contrary to our findings. The observed contraction along the [010] direction leads to an overall lattice expansion in the (010) plane due to Poisson's ratio and agrees well with Fig. 1(c) showing tensile remanent photostriction. The light is therefore able to induce anisotropic deformation in BFO that can be used to stress the magnetostrictive overlayer, as in piezoelectric-magnetostictive structures. This possibility is demonstrated by the experiment in Fig. 2(a), where the 11 nm thick Ni film was deposited on the flat side of the BFO crystal in an $e$-beam evaporator at a rate of 0.1 nm/s for $M(H)$ loop measurements [Fig. 2(a)]. The remanent photostriction largely modifies the magnetic properties of the Ni thin film [Fig. 2(b)], as revealed by the longitudinal magneto-optic Kerr effect (MOKE) magnetometry. The shape of the initial $M(H)$ loop is modified after light exposure, with a change in coercivity of 75%, which remains stable over a long period. For this particular sample, we waited 5 days before electrical recovery tests, but other samples showed that the effect persisted for more than a month. The scenario explaining how light can impact magnetic properties is clearly seen from Fig. 1(a). When the light is turned on, the concentration of free carriers (electrons and holes) starts to increase due to the above-band gap PV effect, and the photocurrent across the BFO crystal stabilizes. This creates an electric field in the bulk of the crystal that tends to influence the net polarization [Fig. 1(a)]. Since the magnitude of this light-induced electric field is small compared to the ferroelectric coercive field, there is no polarization reversal but only slight displacements of the ferroelastic domains in BFO which contribute to its net deformation. After the light is turned off, the generation of free carriers ceases and the ferroelastic domains gradually relax to a new equilibrium configuration that determines the remanent photostriction. This optically induced strain is imprinted in the magnetostrictive Ni adlayer.

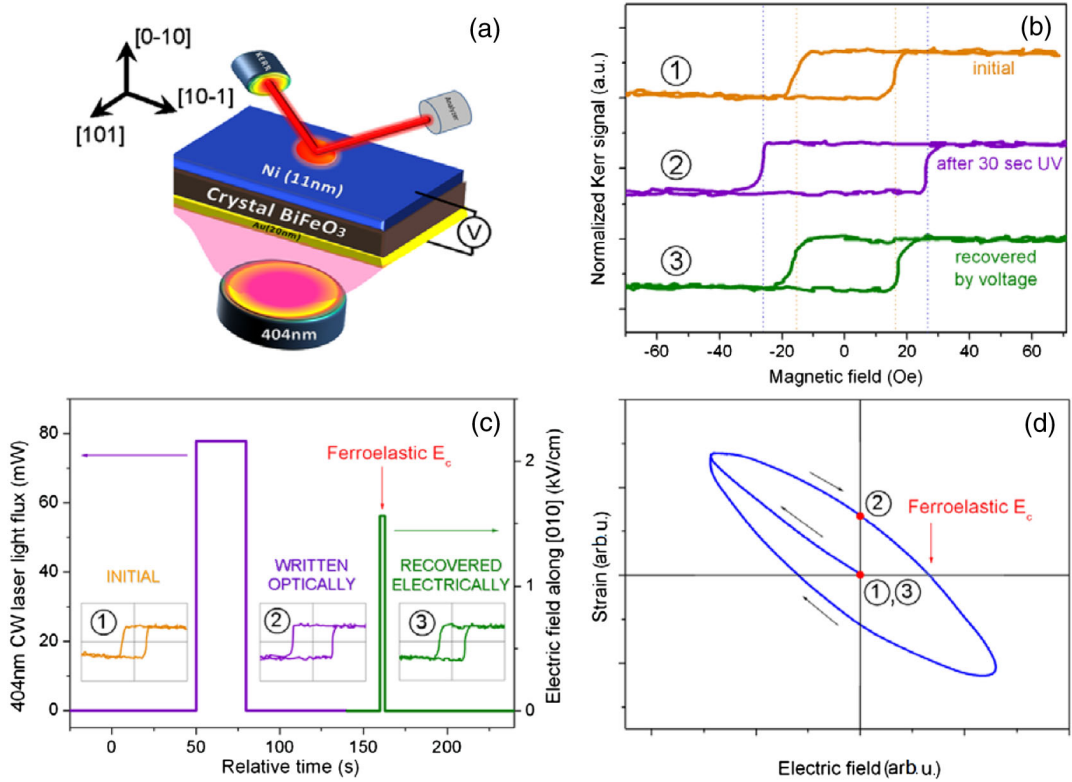

FIG. 2. (a) Schematics of the experiment. (b) Room-temperature ferromagnetic loops of an 11 nm thick Ni film on top of a BFO single crystal before (1) and after (2) excitation by 404 nm light (fluence 250 J cm$^{-2}$). The initial $M(H)$ loop (1) can be recovered (3) by an electric pulse (c) that corresponds to the ferroelastic coercive force $E_c$ as represented by an example sketch (d) [46].

Successful electrical erasing, namely, recovery of the initial ferroelastic configuration of BFO, can be achieved in two ways. If the coercive ferroelastic force is known, it can be done by applying the voltage corresponding to the ferroelastic coercive force [Fig. 2(d)]. The electric field amplitude of 5V/32 μm was enough to recover a close to initial "virgin" $M(H)$ loop in the sample (Fig. 2). Alternatively, an oscillating damped voltage procedure analogous to ac demagnetization can be used, as in the case of electrically written states [17,18]. When the initial spontaneous ferroelastic state is not characterized, the ac electrical erasure may be more convenient.

The possibility of direct ME coupling at the interface [47] can be discarded because the optical writing [Fig. 2(a)] was also demonstrated for samples where a 5 nm Au film is inserted between the BFO substrate and the Ni film to screen any electric charges at the interface. The Au film also excludes the possibility of direct magnetic coupling between the BFO and the Ni.

All MOKE loop measurements were performed at room temperature after excitation and are therefore free of Joule heating artifacts. The data shown in Fig. 1(a) obtained during excitation suggest a negligible heating effect of the laser light, because the polarization of BFO should decrease when warming to its ferroelectric Curie temperature of ~1143 K [48]. A temperature increase of 11.4 K, detected with a thermal camera during the 30 s illumination had no influence on the $M(H)$ loops of the Ni film. Even after heating to 325 K (16 K more than detected by the thermal camera), the $M(H)$ loops remained unchanged. We can therefore safely infer that the optical modification of the magnetic properties has a photovoltaic-photostrictive origin, as confirmed by the electrical erasure test we performed. Our data indicate that the magnetostriction of the Ni adlayer explains the modification of its magnetic properties, originating from the remanent strain state imprinted by light on the BFO substrate.

In conclusion, we have demonstrated that ferroelastic deformation states can be written optically in BFO, and that it is possible to erase them electrically. The remanent photostriction naturally depends on the remanent ferroelectric state of the sample. The possibility to recover the initial state of the functional materials is of key importance, as we observed that the ferroic electric or elastic orders results in remanent states values that depend on the sample's history (spontaneous polarization). This observation requires a special care when performing repetitive experiments (e.g., pump and probe procedures) with unsaturated FE samples in order to guarantee proper reset of the initial polarization. The observed photopolarization induces a deformation that can be coupled to a ferromagnetic adlayer, resulting in optically controlled magnetic anisotropy. This optically induced effect manifests itself in a 75% change in the ferromagnetic coercivity, exceeding by 55% the well-known electric control in the BaTiO$_3$/Fe structures [49] with the nonvolatile and wireless advantage, thus opening the technologically interesting possibility of multistate magnetic operation [Fig. 1(a)]. The ultrafast photostriction in BFO films [50–52] and ceramics [53] combined with the possibility of ultrafast gating [54], provides a perspective for light-controlled magnetic switching devices and magnetoresistive memories on sub-ns time scales. Furthermore, the fact that photostriction can exist in a number of different materials [32,55] expands the horizon of photo-magneto-elastic interactions beyond inorganic compounds [56].

This work was partially supported by French National Research Agency via hvSTRICTSPIN ANR-13-JS 04-0008-01, Labex NIE 11-LABX-0058-NIE and the ANR-10-IDEX-0002-02 research grants. D. S. acknowledges the Helmholtz Association for funding via the Helmholtz Postdoc Programme PD-142. The technical support of the STnano cleanroom (IPCMS) facility is acknowledged. B. K. is grateful to J. M. D. Coey for comments on the manuscript.